\documentclass[pre,twocolumn,aps,superscriptaddress,showpacs,floatfix]{revtex4}

\usepackage{graphicx}
\usepackage{dcolumn}
\usepackage{bm}
\usepackage{amsmath}

\begin{document}
\title{Asymmetry in shape causing absolute negative mobility  }

\author{Peter H\"anggi}
\affiliation{Institut f\"ur Physik, Universit\"at Augsburg, D-86159 Augsburg, Germany}
\author{Fabio Marchesoni}
\affiliation{Dipartimento di Fisica, Universit\`a di Camerino,
I-62032 Camerino, Italy}
\author{Sergey Savel'ev}
\affiliation{Department of Physics, Loughborough University,
Loughborough LE11 3TU, United Kingdom}
\author{Gerhard Schmid}
\affiliation{Institut f\"ur Physik, Universit\"at Augsburg, D-86159 Augsburg, Germany}
\begin{abstract}
We propose a simple classical concept of nanodevices working in an absolute
negative mobility (ANM) regime: The minimal spatial asymmetry required for
ANM to occur is embedded in the geometry of the transported particle, rather
than in the channel design. This allows for a tremendous simplification of device
engineering, thus paving  the way towards practical implementations of ANM.
Operating conditions and performance of our model device are investigated,
both numerically and analytically.
\end{abstract}


 \maketitle

\section{Introduction}\label{intro}

Realizing a micro- or nano-device exhibiting {\it absolute negative mobility} (ANM) poses serious technological
 challenges, as this task is believed to require finely tailored spatial asymmetries either in the (nonlinear)
  particle-particle interactions \cite{INT1,INT2} or, more conveniently, in the geometry of the device itself
\cite{augsburg1a, augsburg1b}. A device is said to operate in the
ANM regime, when it works steadily against a biased force, i.e., a
force with nonzero stationary mean. According to the Second Law of
Thermodynamics (or more precisely, the so called principle of Le
Chatelier), a static force alone cannot induce ANM in a device
coupled to an equilibrium heat bath, unless an additional time
dependent force is applied to bring the system out of equilibrium.
ANM is known to occur as a genuine  {\it quantum mechanical}
phenomenon in photovoltaic materials, as the result of
photo-assisted tunneling in either the bulk of noncentrosymmetric
crystals \cite{quantumANM} or artificial semiconductor structures
\cite{gossarda,gossardb}. However, such manifestations of the ANM
phenomenon do not survive in the limit of a classical description,
so that detecting ANM in a purely classical system remains a
challenging task.

As spatial symmetry typically suppresses ANM, {\it ad hoc} contrived
geometries have been proposed to circumvent this difficulty. The
most promising solution devised to date, is represented by two (2D)
or three dimensional (3D) channels with inner walls tailored so as
to force the transported particles along meandering paths
\cite{augsburg1a,augsburg1b}, a design that can be implemented,
e.g., in superconducting vortex devices \cite{RMP09}. Other
classical set-ups advocate elusive dynamic chaotic effects
\cite{augsburg2a,augsburg2b,bielefeld}. Although such finely tuned
asymmetric geometries and/or nonlinear dynamic behaviors may seem
hardly accessible to table-top experiments, first convincing
demonstrations of classical ANM have actually been obtained
following this strategy \cite{expANMa,expANMb}.

We propose here a much simpler, affordable working concept for a
classical ANM device, by embedding the spatial asymmetry into
the shape of the transported particles, rather than in the channel
geometry. In view of this new formulation, the ANM mechanism is
expected to occur in natural systems, too, where cylindrically
symmetric channels in low spatial dimensions and elongated particles
are frequently encountered \cite{natural,zeolite}.

This paper is organized as follows. We introduce in Sec.
\ref{model} the Langevin equations for a floating ellipsoidal
Brownian particle ac-driven along a 2D compartmentalized channel.
By numerical simulation we show in Sec. \ref{mechanism} that
ANM actually occurs as an effect of the particle elongation.
In Sec. \ref{optim} we analyze the dependence of ANM on both
the drive parameters and the particle geometry, with the purpose
of determining the optimal operating conditions of our model device.
Finally, in Sec. \ref{remarks} we discuss the applicability of the proposed
ANM mechanism to nanoparticle transport in realistic biological and artificial devices.

\begin{figure}[bp]
\centering
\includegraphics[width=9cm,clip]{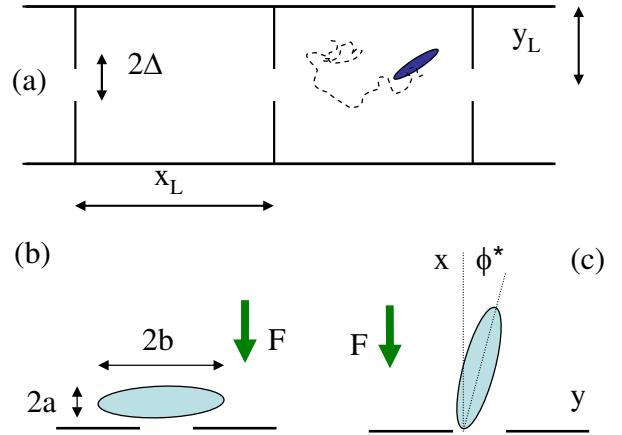}
\caption{(Color online) (a) Asymmetric particle tumbling  in a  periodically segmented 2D-channel.
The pores, $2\Delta$ wide, are centered on the channel axis.
(b) Elliptic particle with semiaxes $a$ and $b$, at rest against
a compartment wall. Note that $a<\Delta<b$. (c) Elliptic particle
in escape position, its major axis forming a maximum angle $\phi^*$ with the channel axis.
\label{F1}}
\end{figure}

\section{The model}\label{model}

Let us consider an elongated Brownian particle, shaped as an elliptic disk,
moving in a straight 2D channel (Fig. \ref{F1}). The overdamped dynamics
of the particle is modeled by three Langevin equations, namely
\begin{subequations}
\label{le}
\begin{align}
\frac{d{\vec r}}{dt}& = -F(t)\;{\vec e}_x + \sqrt{D_r}~{\vec \xi}(t)\,
; \\
\frac{d\phi}{dt} &= \sqrt{D_\phi}~\xi_\phi(t)\, ,
\end{align}
\end{subequations}
where  ${\vec e}_x, {\vec e}_y$ are the unit vectors along the $x,
y$ axes, ${\vec r}\equiv (x,y)$ denotes the particle center of mass,
and $\phi$ is the orientation of its major axis with respect to the
channel axis, ${\vec e}_x$. Here, ${\vec \xi}(t)\equiv
(\xi_x(t),\xi_y(t))$ and $\xi_\phi(t)$ are zero-mean, white Gaussian
noises with autocorrelation functions $\langle \xi_i(t)\xi_j(t')
\rangle = 2\delta_{ij}\delta(t-t')$ and $i,j=x,y,\phi$. The channel
is periodically segmented by means of orthogonal compartment walls,
each bearing an opening, or pore, of half-width $\Delta$, placed at
its center \cite{bu}. As sketched in Fig.~\ref{F1}(a), the channel
is mirror symmetric with respect to both its longitudinal axis and
each compartment wall. This is an important difference with Ref.
\cite{augsburg1a,augsburg1b}, where the channel confining potential,
$V(x,y)$, was taken to be asymmetric under both mirror reflections
-- although symmetric under double reflection, $V(-x,y)\doteq
V(x,-y)$.

In order to detect ANM, the particle must be driven in a pulsating
manner parallel to the channel axis. This means that $F(t)$ consists
of at least two terms \cite{augsburg1a,augsburg1b}: a dc drive,
$F_0$, and an unbiased, symmetric ac drive, $F_{ac}(t)$, with
amplitude $\max \{|F_{ac}(t)|\}= F_1$ and temporal period
$T_\Omega$. Accordingly, the waveform of $F_{ac}(t)$ is subjected to
the conditions $\langle F^{(2n+1)}_{ac}(t)\rangle_\Omega=0$, with
$n=0,1,2 \dots$ and $\langle \dots\rangle_\Omega$ denoting the time
average taken over one drive cycle \cite{criteria}.

Equations (\ref{le}) have been numerically integrated for an elliptic disk of
semiaxes $a$ and $b$, under the assumption that the channel walls were
perfectly reflecting  and the particle-wall collisions were elastic \cite{lboro1}.
In the following we report the outcome of extensive simulations for a fixed channel
compartment geometry, $x_L=y_L$, $\Delta/y_L\ll 1$, but different particle elongations,
$b/a$, ac drive waveforms, and ratios of the rotational to translational diffusion coefficients,
$D_\phi/D_r$. We conclude that ANM occurs in such a highly symmetric channel geometry only because
of the elongated aspect ratio of the drifting particle.

As illustrated in panels (b) and (c) of Fig. \ref{F1}, an elliptic disk with $a<\Delta<b$
crosses a narrow pore only when its major semiaxis forms a small angle with the channel
axis, $|\phi|\leq \phi^*$. For $a\ll b$ (rod-like particle) as in most of our
simulations, $\sin \phi^*\simeq \Delta/b$. To overcome the escape angle $\phi^*$
from a rest position with $\phi=\pi/2$, the disk must rotate against the total
applied force, $F(t)= F_0 + F_{ac}(t)$. For $F_0<F_1$ this is more easily
achieved for pore crossings occurring in the direction of $F_{ac}(t)$,
but opposite to the static force $F_0$. As a result, under appropriate
conditions, detailed below, the net particle current, $\langle v \rangle$,
may indeed flow in the direction {\it opposite} to $F_0$.

\begin{figure}[tp]
\centering
\includegraphics[width=9cm,clip]{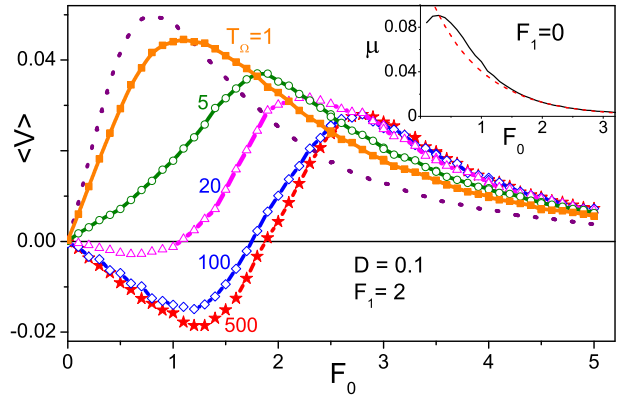}
\caption{(Color online) ANM for a driven-pulsated elongated Brownian particle:
current $\langle v \rangle$ vs. static bias $F_0$ in the presence of
square-wave drives with amplitude $F_1=2$ and different periods $T_\Omega$
(see legend). Other simulation parameters: Diffusion strengths $D=D_r=D_\phi=0.1$;
shape parameters $a=0.05$, $b=0.3$; compartment parameters $x_L=y_L=1$, and $\Delta=0.1$.
Each data point for $\langle v \rangle\equiv \lim_{t\to \infty}\langle x(t)-x(0)\rangle/t$
was computed from a single trajectory with $t=10^6$ and time-step $10^{-5}$; the statistical
error was estimated to be $5\%$, i.e., of the order of the symbol size.
Note, for a comparison, that the compartment traversal time is $\tau_0=7$ and the diffusive
relaxation times are $\tau_D^{(x)}= \tau_D^{(y)}=5$. The dotted curve represents $v(F)$ at
zero ac-drive, 
$F_1=0$. Inset: the corresponding mobility curve (solid curve), $\mu(F)$, is compared with
the analytical estimate of Eq. (\ref{mobility}) (dashed curve) for $F_m=0.86$.
\label{F2}}
\end{figure}

\section{The ANM mechanism}\label{mechanism}

In order to explain the appearance of  ANM, we start looking at the mobility of
an elongated particle driven by a constant force $F$ (Fig. \ref{F2}, inset).
Let $v(F)$ denote its steady velocity and $\mu(F)=v(F)/F$ the relevant
mobility, with $\mu(-F)=\mu(F)$. From now on, and until stated otherwise,
we set for simplicity $D_r=D_\phi=D$, so that $\mu$ is a function of $F/D$.
At equilibrium with $F=0$, $\mu_0 \equiv \mu(0)$ is a $D$-independent constant,
which strongly depends on both the compartment and the particle geometry as
discussed below.
For zero drive, the particle  rotates away from the walls;
pore crossing is thus controlled mostly by {\it translational} diffusion.

As the magnitude of the applied force is increased, the mobility of elliptic and
circular disks develop a quite different $F$ dependence. The mobility of a circular
disk with $a=b<\Delta$, $\mu(F)$, is a concave function of $F/D$, which decays
from $\mu_0\equiv \mu(0)$ to $\mu_\infty \equiv \mu(F\to \infty)=(\Delta-a)/(y_L-a)$,
with a power law slower than $F^{-1}$ \cite{lboro1}. In the case of an elliptic disk with $a<\Delta<b$,
reaching the escape angle $\phi^*$ can be regarded as a noise activated process with energy barrier
proportional to $F$. 
Pore crossing will then be controlled mostly by the {\it rotational} fluctuations, with
approximate escape time
\begin{equation} \label{time}
\tau_0(F)=\tau_0 \exp{(F/F_m)},
\end{equation}
where $F_m=2D/(b\cos \phi^*-a)$ is the total activation force, the factor 2 accounts for the two
directions of rotation, and $\tau_0$ is the compartment traversal time, $x_L/F$, divided by the probability,
$p=(\Delta-a)/(y_L-a)$, that the disk slides through the pore without an additional rotation.
The reciprocal of $\tau_0$ plays the role of an effective attack frequency. This estimate for the
particle crossing time surely holds good for $b\gg \Delta$ and $F \gg F_m$, where
\begin{align}
\tau_0 \ll \tau^{(x)}_D,\tau^{(y)}_D \ll \tau_0(F)\, ,
\end{align}
 with $\tau^{(x)}_D=x_L^2/2D$ and $\tau^{(y)}_D=y_L^2/2D$
denoting, respectively, the longitudinal and transverse relaxation times.

The curve $\mu(F)$ for an elongated particle is thus concave for $F < D/x_L$
\cite{lboro1} and decays exponentially for $F\gg F_m$, like
\begin{equation} \label{mobility}
\mu(F)\simeq \frac{x_L}{F\tau_0(F)} = 2p\exp{(-F/F_m)},
\end{equation}
see inset in Fig. \ref{F2}. Correspondingly, $v(F)$ increases like $\mu_0F$
at small $F$ and decays to zero like $x_L/\tau(F)$ at large $F$, going through a maximum
for $F \sim F_m$, as confirmed by our simulations, see Fig. \ref{F2}.

\begin{figure}[tp]
\centering
\includegraphics[width=8cm,clip]{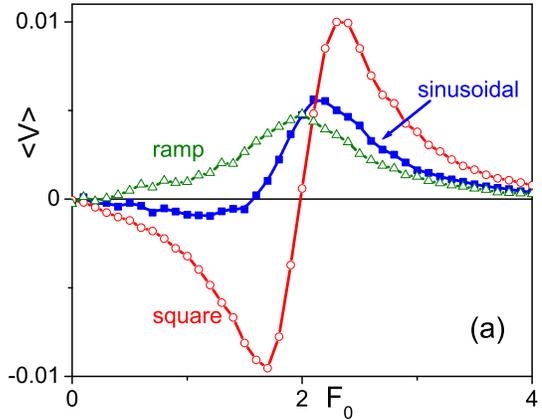}
\includegraphics[width=8cm,clip]{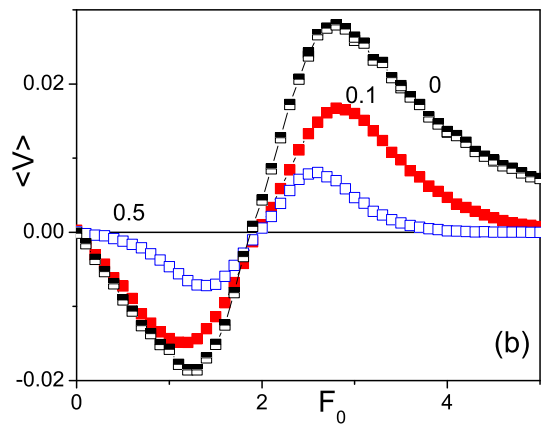}
\caption{(Color online) (a) Drive waveform dependence: $\langle v \rangle$ vs.
$F_0$ for three waveforms of $F_{ac}(t)$, square (as in Fig. \ref{F2}),
sinusoidal and ramped, all with $F_1=2$, $D=0.02$, and $T_\Omega=10^3$.
(b) Particle inertia dependence: $\langle v \rangle$ vs. $F_0$ for three
values of $m=I$, see text. Other simulation parameters are $F_1=2$, $D=0.1$, and $T_\Omega=500$.
\label{F3}}
\end{figure}

A convincing evidence for ANM has been obtained by simultaneously applying to the elliptic
disk a tunable dc force, $F_0$, and a low-frequency, square-wave ac force, $F_{ac}(t)$, with
amplitude $F_1>F_0\geq 0$. The characteristics curves $\langle v\rangle$
{\it vs.} $F_0$ plotted in Fig. \ref{F2}  exhibit a negative ANM branch only for sufficiently
long ac drive periods and $F_0<F_1$; for $F_0>F_1$, however, the current is always oriented
in the $F_0$ direction, no matter what 
$T_\Omega$.

This behavior can be
explained in the {\it adiabatic} regime, where a half drive
period, $T_{\Omega}/2$, is larger than all the drift and diffusion times inside a channel
compartment, namely, $\tau_{0}$, $\tau^{(x)}_D$, and $\tau^{(y)}_D$ \cite{chemphyschem}.
Note that in the opposite regime, ANM is suppressed. The net current can then be approximated by
\begin{equation}\label{vsquare}
\langle v(F_0)\rangle = \frac{1}{2} [v(F_1+F_0)-v(F_1-F_0)].
\end{equation}
As the curve $v(F)$ peaks at $F\sim F_m$, we expect $\langle v(F_0)\rangle$ to develop a negative
minimum for $F_0=F_1-F_m$ and a positive maximum for $F_0=F_1+F_m$, as shown in Fig. \ref{F2}.
As this holds true only for $F_1 > F_m$, the two peaks have upper bounds
\begin{align}
  |\langle v(F_1-F_m)\rangle| \lesssim \langle v(F_1+F_m)\rangle
  \lesssim v(F_m)\, .
\end{align}

For a more quantitative analysis of this phenomenon, we rewrite $v(F_1\pm F_0)=(F_1\pm F_0)
\mu(F_1\pm F_0)$, so that the ANM condition, $\langle v\rangle<0$, reads
\begin{equation} \label{ANM}
\frac{F_1 - F_0}{F_1 +F_0} > \frac{\mu(F_1 +F_0)}{\mu(F_1 -F_0)}\simeq e^{-2F_0/F_m}.
\end{equation}
The approximate equality on the r.h.s. applies for $0<F_0<F_1-F_m$, where
\begin{equation}
\mu(F_1 \pm F_0)\simeq \mu(F_1) e^{\mp F_0/F_m}.
\end{equation}

The above inequality is satisfied for $0<F_0<F^*$, with the turning point, $F^*$,
shifting towards zero in the limit $F_1 \to F_m+$, and towards $F_1$ in the opposite limit,
 $F_1 \gg F_m$. The approximate equality in Eq. (\ref{ANM}) leads to slightly overestimating
  $F^*$, with no prejudice of our conclusion: In the adiabatic regime, ANM  occurs in an appropriate
  $F_0$ interval $(0,F^*)$ only provided that $F_1>F_m$.

\begin{figure}[tp]
\centering
\includegraphics[width=9cm,clip]{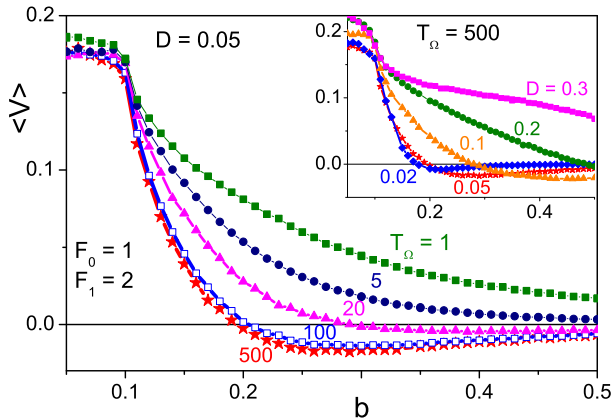}
\caption{(Color online) Particle elongation dependence: $\langle v \rangle$ {\it vs.}
 $b$ for different values of $T_\Omega$ (main panel) and of $D$ (inset).
 Other simulation parameters: square-wave ac-drive with amplitude strength
 $F_1=2$, and bias $F_0=1$, $a=0.05$, $x_L=y_L=1$, and $\Delta=0.1$. \label{F4}}
\end{figure}

Finally, we notice that for $F_0>F_1>F_m$ the net current reads
\begin{align}
  \langle v(F_0)\rangle = \frac{1}{2}[v(F_0+F_1)+v(F_0-F_1)]
\end{align}
and for extremely large $F_0$, it decays to zero like
$\langle v(F_0)\rangle \sim \frac{1}{2}v(F_0-F_1)$.
Correspondingly, our simulation data in the
neighborhood of the turning point $F^* \sim F_1$ are reasonably well
reproduced by the linear fitting law, $\langle v(F_0)\rangle \sim (\mu_0/2)(F_1-F_0)$.

\section{Selectivity and Optimization criteria}\label{optim}

In view of future experimental implementations of the proposed ANM mechanism, we now analyze
in detail its sensitivity with respect to both the drive and the particle parameters.

We start noticing that the most prominent ANM effect is produced, in fact,
by the square waveform $F_{ac}(t)$ adopted in Fig. \ref{F2}. For the sake of a
comparison, in Fig. \ref{F3}(a) we plotted $\langle v(F_0)\rangle$ also for other, inversion-symmetric
waveforms $F_{ac}(t)$  with the same amplitude and period, in particular, sinusoidal
and up-down ramped waveforms. For a ramped ac drive, no ANM can occur, because in the adiabatic regime
\begin{align}
  \langle v(F_0)\rangle=\frac{1}{2F_1}\int_{F_1-F_0}^{F_1+F_0}v(F)dF
  \geq 0\, .
\end{align}
For a sinusoidal ac drive, the ANM effect can be shown analytically to diminish in
magnitude and shrink to a narrower interval $(0,F^*)$ than obtained for the corresponding square
waveform. The latter is thus the optimal ac drive waveform to operate an ANM device.

To quantify the robustness of this effect against the damping conditions, inertia was
added to the model by replacing the l.h.s. in the Langevin equations (\ref{le}) as follows:
\begin{subequations}
  \label{inertia}
\begin{align}
\frac{d{\vec r}}{dt}&\to  \frac{d{\vec r}}{dt}-m\frac{d^2{\vec r}}{dt^2}\\
\frac{d\phi}{dt} & \to \frac{d\phi}{dt}-I\frac{d^2\phi}{d^2t}\, .
\end{align}
\end{subequations}
In Fig. \ref{F3}(b) we compare ANM characteristics curves for growing values of the
(rescaled) particle mass, $m$, and moment of inertia, $I$: ANM is gradually suppressed
by increasing inertia. This is no serious limitation, as in most experiments rectifiers
operate, indeed, under overdamped, or zero mass, conditions \cite{RMP09}.

We analyze next how selective the ANM effect is versus the geometric and diffusive
properties of the transported particles. In Fig. \ref{F4} we displayed the dependence
of the net current on the particle elongation. One notices immediately that, when
plotted versus $b$ at constant values of the drive parameters,
$\langle v \rangle$ starts out positive and then turns negative for $b$ larger than
a certain threshold, $b^*$, which appears to increase with either raising $D$
(figure inset) or lowering $T_\Omega$ (main panel).

Our adiabatic argument provides a simple explanation for these findings, as well.
We recall that the mobility curve $\mu(F)$ decays exponentially on a scale
$F_m\propto b^{-1}$. As a consequence, for $b \to \infty$, Eq. (\ref{vsquare})
boils down to $\langle v \rangle \simeq -\frac{1}{2}v(F_1-F_0)$,
which means that $\langle v \rangle$ tends to zero from negative values,
in agreement with our data. For $b\leq \Delta$ the particle flows through the pores,
no matter what the orientation, $\phi$, of its major axis. Therefore, $\langle v \rangle$
becomes insensitive to the particle elongation (main panel), while retaining its known $D$ dependence (inset).

The actual value of $b^*$ is determined by the general ANM condition (\ref{ANM}).
On making use of the approximation on the r.h.s. of that equation, one easily proves
the existence of the threshold $b^*$ for any geometry and drive parameter set.
We caution that this way one may underestimate $b^*$ and, therefore,
the predicted dependence $b^* \propto D$ holds qualitatively, only
(see inset of Fig. \ref{F4}). Of course, when $D$ is raised so that $b^*$
grows larger than the spatial dimensions of a channel compartment, then ANM is suppressed altogether.

\begin{figure}[tp]
\centering
\includegraphics[width=8.5cm,clip]{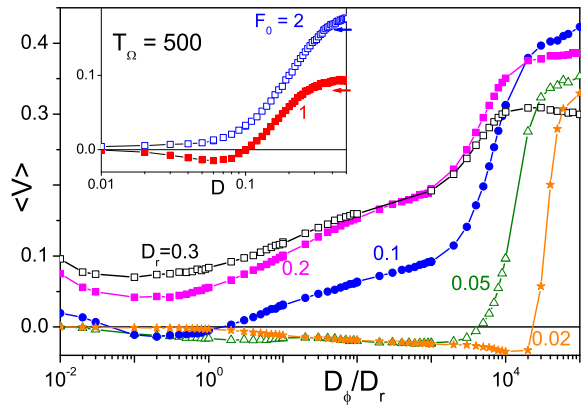}
\caption{(Color online) Rotational translational diffusion dependence: $\langle v \rangle$ {\it vs.}
 $D_\phi/D_r$ for different values of $D_r$ (see legend). Inset:  $\langle v \rangle$ {\it vs.}
  $D=D_r=D_{\phi}$
with $D_\phi/D_r=1$. The arrows mark the asymptotes predicted in the text.
Other simulation parameters are as in Fig. \ref{F4} with $b=0.3$.}
\label{F5}
\end{figure}

The dependence of the current on the fluctuation intensities is illustrated in Fig. \ref{F5}.
We consider first the case $D_r=D_\phi=D$ (figure inset). The dependence of $\langle v \rangle$ on $D$
can be analyzed following the approach introduced to interpret the results in Fig. \ref{F4}.
On recalling that $F_m \propto D$, in the limit $D \to \infty$, the ac drive amplitude ends
up being smaller than $F_m$, $F_m>F_1$, thus suppressing ANM. In the opposite limit,
$D\to 0$, $F_m$ vanishes and ANM is predicted to occur for any dc drive such that
$0<F_0<F_1$. Indeed, from Eq. (\ref{vsquare}) we obtain $\mu(F_1\pm F_0) \to \mu_0$
or $\langle v \rangle \to \mu_0 F_0>0$, for $D\to \infty$ (marked in figure by
horizontal arrows), and $\langle v \rangle \simeq -\frac{1}{2}v(F_1-F_0) \to 0-$,
for $D \to 0$. On using $D$ as a control parameter, ANM is thus restricted to
low noise, $0<D<D^*$, with the threshold $D^*$ also obtainable from the ANM condition (\ref{ANM}).

We consider next the more general case when $D_r$ and $D_\phi$ can be independently varied,
while keeping $a$ and $b$ fixed. In Fig. \ref{F5}, $\langle v \rangle$ has been plotted versus
$D_\phi/D_r$ for different values of $D_r$. The two opposite limits of the net current,
$\langle v \rangle_0$, for  $D_\phi/D_r\to 0$, and $\langle v \rangle_\infty$, for
$D_\phi/D_r\to \infty$, are both positive with $\langle v \rangle_0<\langle v \rangle_\infty$.
In between, the magnitude of the ANM effect is seemingly not much
sensitive to $D_\phi/D_r$ over several orders of magnitude, which
allows us to generalize the conclusions drawn above for
$D_\phi/D_r=1$ to the case of realistic extended particles. Note
that $\langle v \rangle_0$ is non-null because an elliptic disk with
$a<\Delta<b$ can diffuse across a pore even in the limit $D_\phi\to
0$, where the adiabatic argument fails, thanks to the sole
translational fluctuations. In view of the crossing condition
$|\phi|<\phi^*$, an elongated particle can be handled as a circular
one with radius smaller that $\Delta$, see Fig. \ref{F4}, but
crossing probability $2\phi^*/\pi$. This argument can be extended to
the case of $\langle v \rangle_\infty$, with the important
difference that for
$D_\phi\to \infty$ the particle has crossing
probability one. Both $\langle v \rangle_0$ and $\langle v
\rangle_\infty$ are thus positive, with $\langle v \rangle_0$
relatively smaller than $\langle v \rangle_\infty$.

\section{Concluding remarks}\label{remarks}

The ANM model presented in this work, although stylized, lends itself to interesting nano-technological
applications \cite{RMP09,chemphyschem}. A typical reference case is represented by the transport of hydrated DNA
fragments across narrow compartmentalized channels \cite{dekker}. Rod-like DNA fragments $30$-$40$nm in length
and $1$-$2$nm in (hydrated) diameter are easily accessible \cite{tirado}; their elongation ratio is about $3$
 times larger than $b/a$ in Fig. \ref{F2}, but still within the ANM range of Fig. \ref{F4}. Artificial
 nanopores can be TEM drilled in $10$nm thin SiO$_2$ membranes with reproducible diameters of $5$nm,
 or less \cite{dekker}, which is consistent with the elongation selectivity condition $a<\Delta<b$
 assumed throughout this work. Moreover, the measured $D_\phi/D_r$ ratio for the hydrated DNA
 fragments of Ref. \cite{tirado} falls in the range $100$-$200$, in
dimensionless units, where ANM can also occur, as shown in Fig. \ref{F5}, for an appropriate
choice of the drive parameters. To this regard, we remind that experiments on
DNA translocation across artificial nanopores require applied electrical fields of
the order of 10-100kV/cm \cite{healy}; if applied to the DNA rods of Ref. \cite{tirado},
electrical fields of that intensity, or less, would satisfy the ANM condition of Eq.
(\ref{ANM}) with $F_1>F_m$ at room temperature.

The ANM characteristics curves plotted in Figs. \ref{F2} and \ref{F5}, being quite selective with respect to
the particle shape, suggest the possibility of developing artificial devices that efficiently operate
as {\it geometric sieves} for nanoparticles.

Our model was stylized to capture the key mechanism responsible for
the occurrence of ANM in symmetric channels. The mechanism
summarized by Eqs. (\ref{time}) and (\ref{ANM}), however, clearly
does not depend on the dimensionality of the channel (experiments
can then be carries out in 3D geometries), but can be impacted by
other competing effects: (i) Pore selectivity. For a given
translocating molecule, the actual crossing time varies with the
wall structure inside the pore and in the vicinity of its opening
\cite{iqbal}; (ii) Electrophoretic effects. The inhomogeneous
electrical field generated by the electrolyte flow across the pores
acts on the orientation of drifting spheroidal particles
\cite{solomontsev}. System specific effects (i) and (ii) can readily
be incorporated in our model by adding appropriate potential terms,
$U_r(x,y)$ and $U_\phi(\phi)$, to the Langevin equations (\ref{le}).

\acknowledgements{
The work is supported by the Humboldt prize program (F.M.), Humboldt-Bessel prize program
(S.S.), the Volkswagen foundation (P.H., G.S.), project I/83902, The European Science Foundation
(ESF) under its  program ``Exploring the physics of small devices" (P.H.) and by the German Excellence
Initiative via the Nanosystems Initiative Munich (NIM) (P.H., G.S.).}


\begin{references}
%
\bibitem{INT1}P. Reimann, R. Kawai, C. Van den Broeck, and P. H\"anggi, Europhys. Lett. {\bf 45}, 545 (1999).
%
\bibitem{INT2}B. Cleuren and C. Van den Broeck, Europhys. Lett. {\bf 54}, 1 (2001).
%
\bibitem{augsburg1a} R. Eichhorn, P. Reimannn, and P. H\"anggi, Phys. Rev. Lett. {\bf 88}, 190601 (2002).;
%
\bibitem{augsburg1b} R. Eichhorn, P. Reimannn, and P. H\"anggi, Phys. Rev. E {\bf66}, 066132 (2002).
%
%
\bibitem{quantumANM} B.I. Sturman and V.M. Fridkin, {\it The Photovoltaic and Photorefractive
Effects in Noncentrosymmetric Materials} (Gordon and Breach Science Publishers, Philadelphia, 1992).
%
\bibitem{gossarda} R.A. H\"opfel J. Shah, P.A. Wolff, and A.C. Gossard, Phys. Rev. Lett.
{\bf 56}, 2736 (1986).
\bibitem{gossardb} B.J. Keay, S. Zeuner, S.J. Allen, K.D.
Maranowski, A.C. Gossard, U. Bhattacharya, and M.J.W. Rodwell, Phys.
Rev. Lett. {\bf 75}, 4102 (1995).
%
\bibitem{RMP09} P. H\"anggi and F. Marchesoni, Rev. Mod. Phys. {\bf 81}, 387 (2009).
%
\bibitem{augsburg2a} L. Machura, M. Kostur, P. Talkner, J. Luczka, and P. H\"anggi, Phys. Rev. Lett.
{\bf 98}, 040601 (2007).
%
\bibitem{augsburg2b} M. Kostur, L. Machura, P. Talkner, P. H\"anggi, and J. Luczka,
 Phys. Rev. B {\bf77}, 104509 (2008).
%
\bibitem{bielefeld} D. Speer, R. Eichhorn, and P. Reimann, Phys. Rev. E {\bf 76}, 051110 (2007).
%
\bibitem{expANMa} A. Ros, R. Eichhorn, J. Regtmeier, T.T. Duong, P. Reimann, and D. Anselmetti,
Nature (London) {\bf 436}, 928 (2005).

\bibitem{expANMb} J. Nagel, D. Speer, T. Gaber, A. Sterck, R. Eichhorn, P. Reimann,
K. Ilin, M. Siegel, D. Koelle, and R. Kleiner, Phys. Rev. Lett. {\bf
100}, 217001 (2008).
%
\bibitem{natural} B. Hille, {\it Ion Channels of Excitable Membranes} (Sinauer, Sunderland, 2001).
%
\bibitem{zeolite} J. K\"arger and D.M. Ruthven, {\it Diffusion in Zeolites and other Microporous Solids}
(Wiley, New York, 1992).
%
\bibitem{bu} M. Borromeo and F. Marchesoni, Chem. Phys. {\bf 375}, 536 (2010).
%
\bibitem{criteria}P. H\"anggi, R. Bartussek, P. Talkner, and J. Luczka, Europhys. Lett. {\bf 35}, 315
(1996).
%
\bibitem{lboro1} F. Marchesoni and S. Savel'ev, Phys. Rev. E {\bf 80}, 011120 (2009).
%
\bibitem{chemphyschem} P.S. Burada, P. H\"anggi, F. Marchesoni, G. Schmid,
and P. Talkner, ChemPhysChem {\bf 10}, 45 (2009).
%
%
%
\bibitem{dekker} C. Dekker, Nature Nanotech. {\bf 2}, 209 (2007).
%
\bibitem{tirado} M.M. Tirado, C.L. Mart\'inez, and J.G. de la Torre, J. Chem. Phys {\bf 81}, 2047 (1984).
%
\bibitem{healy} For a review see: K. Healy,  Nanomedicine {\bf 2}, 459 (2007).
%
\bibitem{iqbal} S.M. Iqbal, D. Akin, and R. Bashir, Nature Nanotech. {\bf 2}, 243 (2007).
%
\bibitem{solomontsev} Y. Solomentsev and J.L. Anderson, Ind. Eng. Chem. Res. {\bf 34}, 3231 (1995).


\end{references}
\end{document}